\documentclass[10pt]{article}
\usepackage{latexsym}
\usepackage{graphicx}
\usepackage{multirow}
\usepackage{amsmath}
\usepackage{ulem}
\usepackage{setspace}
\usepackage{subfigure}
\begin{document}
\begin{center}\Large{Gravitational interactions between fast neutrinos and the formation of bound rotational states}
\end{center}
\begin{center}
\renewcommand\thefootnote{\fnsymbol{footnote}}
Constantinos G. Vayenas$^{1,2,}$ \footnote{E-mail: cgvayenas@upatras.gr} \& Stamatios Souentie$^1$
\end{center}
\begin{center}
\textit{$^1$LCEP, Caratheodory 1, St., University of Patras, Patras GR 26500, Greece\\$^2${Division of Natural Sciences, Academy of Athens, Panepistimiou 28, Ave., GR-10679 Athens, Greece}}
\end{center}
\begin{abstract}
The gravitational forces exerted between fast neutrinos at short distances are examined using Newton's gravitational law, special relativity, and the equivalence principle. It is found that the magnitude of these forces is not negligible and can lead to the formation of bound rotational states with radii in the $fm$ range. 

{\textbf{PACS}: {03.30.+p, 13.15.+g, 14.60.Lm, 12.60.Rc, 14.20.-c}}
\end{abstract}

\section{Introduction}
During the last two decades the study of neutrino oscillations \cite{Fukuda98,Mohapatra07} has shown conclusively that all the three types, or flavors, of neutrinos ($\nu_e,\nu_\mu,\nu_\tau$) have small but non-zero rest masses \cite{Fukuda98,Mohapatra07,Griffiths08}. While their rest masses are very small $(\sim 0.04$ to 0.4 $eV/c^2)$ \cite{Fukuda98,Mohapatra07,Griffiths08} neutrinos have typically quite large (as high as 200 $MeV$) total energies \cite{Fukuda98,Mohapatra07}. This implies that their velocities are very near the speed of light and that their Lorentz factors $\gamma (=(1-\texttt{v}^2/c^2)^{-1/2})$ are very large. Since the total energy, $E$, is related to the rest mass, $m_o$, via the Einstein equation:
\begin{equation}
E=\gamma m_oc^2
\label{eq:1}
\end{equation}
it follows, as an example, that for $E=200\;MeV$ and $m_o=0.04\;eV/c^2$ it is $\gamma m_o=200\;MeV/c^2$ and thus $\gamma=5\cdot 10^9$.

On the other hand, special relativity dictates that when a particle moving on an instantaneous frame $S'$ has a velocity $\textbf{v}$ and concomitant Lorentz factor $\gamma$ relative to a laboratory observer in a frame $S$, then, at least for linear motion, the inertial mass, $m_i$, equals $\gamma^3m_o$ \cite{French68,Freund08}. This is obtained after some simple algebra from \cite{French68,Freund08}:
\begin{equation}
F=\frac{dp}{dt}=\frac{d(\gamma m_o\texttt{v})}{dt}=m_o\left[\gamma+\gamma^3\frac{\texttt{v}^2}{c^2}\right]\frac{d\texttt{v}}{dt}=\gamma^3m_o\frac{d\texttt{v}}{dt}
\label{eq:2}
\end{equation}
where $p$ is the momentum.

However, according to the equivalence principle the inertial mass, $m_i$, of a particle equals the gravitational mass, $m_g$ \cite{Schwarz04,Gine09}, and thus one obtains: 
\begin{equation}
m_g=m_i=\gamma^3m_o
\label{eq:3}
\end{equation}

Upon substituting the above values, i.e. $\gamma=5\cdot 10^9$ and $m_o=0.04\;eV/c^2$, one finds:
\begin{equation}
m_g=m_i=5\cdot 10^{18}\;GeV/c^2
\label{eq:4}
\end{equation}
which, surprisingly, is almost half the value of the Planck mass \cite{Schwarz04}, i.e. $1.221\cdot 10^{19}\;GeV/c^2$ $(=2.177\cdot 10^{-8}\:kg)$. Since the magnitude of gravitational and strong forces are expected to merge at energies close to the Planck energy, $1.221\cdot 10^{19}\;GeV$, it follows that the gravitational forces between such fast neutrinos can be quite significant and thus can lead to the creation of confined states. 

For example the gravitational potential energy, $V_g$, of two such fast moving particles when they are at a distance of 1 $fm$ is:
\begin{equation}
V_g=-\frac{Gm^2_g}{r}=-5.30\cdot 10^{-12}\;J=-33.09\;MeV
\label{eq:5}
\end{equation}
whereas for comparison the Coulombic potential energy of a $u$ and a $d$ quark at the same distance is:
\begin{equation}
V_c=-\frac{(2/3)(1/3)e^2}{\epsilon r}=-5.126\cdot 10^{-14}\:J=-0.32\;MeV
\label{eq:6}
\end{equation}
, i.e. the gravitational interaction is, surprisingly, a factor of 100 stronger than the Coulombic interaction. 

Since the strong force interaction between quarks is estimated to be a factor of $\alpha^{-1}(=137.035)$ stronger than the Coulombic interaction at the $fm$ range \cite{Schwarz04,Hooft07,Povh06}, it follows that the magnitude of the gravitational force between fast neutrinos, when accounting for special relativity and for the equivalence principle, can be comparable to the magnitude of the strong force at the $fm$ range. 

This result is at first surprising but stems directly from special relativity, i.e. eqs (\ref{eq:2}) and (\ref{eq:3}), from the weak equivalence principle of E\"{o}tv\"{o}s and Einstein \cite{Schwarz04} (eq. \ref{eq:3}), and from Newton's gravitational law, without making any assumptions. 

It thus becomes interesting to examine what type of bound states such a powerful attractive force can create.
 
\section{Circular states}
We thus examine the circular motion of three neutrinos (e.g. three electron neutrinos or antineutrinos or a combination thereof) on a circle of radius $R$ (Fig. 1) under the influence of their gravitational attraction.

For circular motion the centripetal force, $\textbf{F}$, is related to the radius, $R$, and velocity, $\textbf{v}$, via:
\begin{equation} 
\label{eq:7} 
F=\gamma m_o \frac{\texttt{v}^2}{R}
\end{equation}

This is derived from the general form of the relativistic equation of motion \cite{French68,Freund08}, i.e.:
\begin{equation}
\label{eq:8}
\textbf{F}=\frac{d\textbf{p}}{dt}=\gamma m_o\frac{d\textbf{v}}{dt}+\gamma^3m_o\frac{1}{c^2}\left(\textbf{v} \bullet\frac{d\textbf{v}}{dt}\right)\textbf{v}
\end{equation}
by accounting for $\textbf{v}\bullet(d\textbf{v}/dt)=0$.

The next step is to express the centripetal force, $\textbf{F}$, in terms of the gravitational force exerted to each particle by the two other particles. This will be done via Newton's universal gravitational law by utilizing the actual gravitational masses of the three particles and not their rest masses, $m_o$, or their relativistic masses, $\gamma m_o$. According to the equivalence principle the gravitational mass equals the inertial mass of each particle. 
\begin{figure}[ht]
\begin{center}
\includegraphics[width=0.35\textwidth]{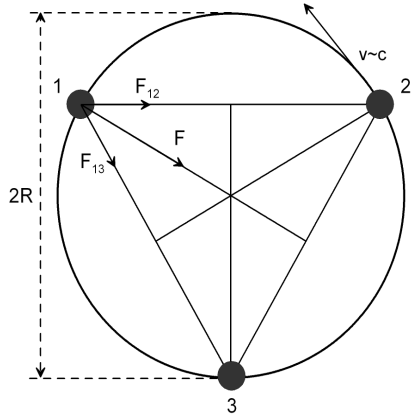}
\end{center}
\caption{Three neutrinos moving at a constant tangential velocity, $\textbf{v}$, in a circle of radius $R$ around their center of mass. They are equally spaced. $F_{12}$ and $F_{13}$ are two particle attraction forces and $F$ is the resultant, radial, force.}
\label{fig:1}
\end{figure}

\subsection{Equivalence principle and inertial mass}
It is thus important to first examine if equation (\ref{eq:3}) for the particle inertial and gravitational mass, obtained via equation (\ref{eq:2}) for linear particle motion, is also applicable when the particle performs a circular motion. 

We thus consider a laboratory frame $S$ and an instantaneous inertial frame $S'$ moving with a particle with an instantaneous velocity  $\textbf{v}$ relative to frame $S$ (Figure 2).
\begin{figure}[ht]
\begin{center}
\includegraphics[width=0.27\textwidth]{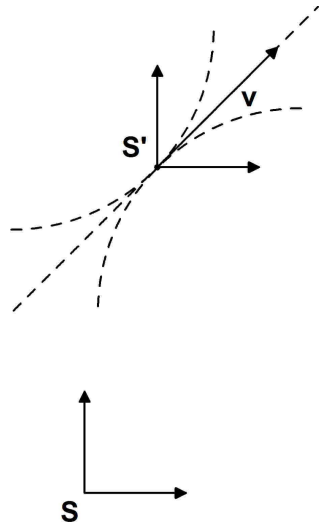}
\caption{Laboratory frame $S$ and instantaneous inertial frame $S'$,the latter moving with the particle under consideration. The frame $S'$ is uniquely defined by the vector $\textbf{v}$ alone regardless
of the motion (e.g. linear or circular) performed by the particle.}
\label{fig:2}
\end{center}
\end{figure}

It is worth noting that the instantaneous inertial frame $S'$ is defined by the vector $\textbf{v}$ alone and not by the overall type of motion (e.g. linear or cyclic) performed by the particle \cite{French68,Freund08}.

For the laboratory observer in $S$ a particle in the instantaneous frame $S'$ performing a circular motion is indistinguishable from a particle of the same rest mass $m_o$ and velocity, $\textbf{v}$, performing a linear motion (Fig. 2).
 
The inertial mass of a particle, $m_i$ is always (e.g. \cite{Gine09}) defined as the ratio of force and acceleration:
\begin{equation}
m_i=\frac{\textbf{F}}{d\textbf{v}/dt}
\label{eq:9}
\end{equation}

According to the weak equivalence principle, which has been confirmed at the level of 1 part of $10^{12}$ \cite{Schwarz04}, the particle inertial mass equals the particle gravitational mass $m_g$:
\begin{equation}
m_i=m_g
\label{eq:10}
\end{equation}

Since $m_i$ is a scalar, it follows from (\ref{eq:9}) that $d\textbf{v}/dt$ and the force $\textbf{F}$ used to compute $m_i$ in (\ref{eq:9}) have to be colinear. Thus one can examine three cases:\\ 
\textbf{A. Colinear $\textbf{F}$ and $\textbf{v}$}: If $\textbf{F}$ and $\textbf{v}$ are colinear then $d\textbf{v}/dt$ is also colinear and the relativistic equation of motion, eq. (\ref{eq:8}): 
\begin{equation*}
\textbf{F}=\frac{d\textbf{p}}{dt}=\gamma m_o\frac{d\textbf{v}}{dt}+\gamma^3m_o\frac{1}{c^2}\left(\textbf{v} \bullet\frac{d\textbf{v}}{dt}\right)\textbf{v}
\end{equation*}
reduces to:
\begin{eqnarray} 
\label{eq:11} 
F=\frac{dp}{dt}&=&\gamma m_o\frac{d\texttt{v}}{dt}+\gamma^3m_o\frac{1}{c^2}\texttt{v}^2\frac{d\texttt{v}}{dt}=
\left[\gamma+\gamma^3\frac{\texttt{v}^2}{c^2}\right]m_o\frac{d\texttt{v}}{dt}\\
\nonumber &=&\left[\gamma+\gamma^3(\gamma^2-1)/\gamma^2\right]m_o\frac{d\texttt{v}}{dt}=\gamma^3m_o\frac{d\texttt{v}}{dt}
\end{eqnarray}
This result is already known from the pioneering 1905 paper of A. Einstein on special relativity [\cite{Einstein1905}, he uses the symbol $\beta$ for $\gamma$]. It defines the mass $\gamma^3m_o$, frequently termed longitudinal mass \cite{Freund08}, which is the inertial mass of the particle, $m_i$, as it equals the ratio of force and acceleration, i.e. eq. (\ref{eq:3}):
\begin{equation*}
m_g=m_i=\gamma^3m_o
\end{equation*}

According to the theory of special relativity \cite{French68} the case where $\textbf{F}$ and $\textbf{v}$ are parallel is the only case where $\textbf{F}$ is \textit{invariant}, i.e. where the same value of $F$ is perceived in the inertial frames $S$ and $S'$ \cite{French68}. It is obvious that only when this is the case (i.e. the force value perceived by the observer at $S$ is the same with that acting on the particle) then a meaningful for the observer use of $\textbf{F}$ in equation (\ref{eq:9}) or (\ref{eq:11}) can be made. 

For example for circular motion $\textbf{F}$ and $\textbf{v}$ are not colinear and thus, although the last term in equation (\ref{eq:8}) vanishes, one cannot assign the value $\gamma m_o$ to the inertial mass. Some paradoxes arising from such an erroneous assignment are given in the Appendix.  

\textbf{B. Zero force:} In this case the inertial mass, $m_i$ must again be found from eq. (\ref{eq:9}): 
\begin{equation*}
m_i=\textbf{F}/(d\textbf{v}/dt)
\end{equation*}
where $\textbf{F}$ and $\textbf{v}$ are colinear \cite{French68}. 

Thus in order to utilize equation (\ref{eq:8}) one can consider that an infinitesimal $\textit{test}$ force $\delta\textbf{F}$, parallel to $\textbf{v}$, is acting on the particle \cite{French68} causing an infinitesimal acceleration $d\delta \textbf{v}/dt$. Since the initial $\textbf{v}$ has been assumed constant, the latter is also equal to $d\textbf{v}/dt$.

We can thus now apply the relativistic equation of motion \cite{Freund08}, eq. (\ref{eq:8}):
\begin{equation*}
\textbf{F}=\frac{d\textbf{p}}{dt}=\gamma m_o\frac{d\textbf{v}}{dt}+\gamma^3m_o\frac{1}{c^2}\left(\textbf{v}\bullet\frac{d\textbf{v}}{dt}\right)\textbf{v}
\end{equation*}
to our case (setting $\textbf{F}=\delta \textbf{F}$ parallel to \textbf{v}), to obtain: 
\begin{eqnarray} 
\label{eq:12}
\delta F&=&\gamma m_o+\gamma^3\frac{m_o}{c^2}\texttt{v}^2\frac{d(\delta \texttt{v})}{dt}=\\
\nonumber &=& m_o\left[\gamma+\gamma^3\frac{\texttt{v}^2}{c^2}\right]\frac{d(\delta \texttt{v})}{dt}\\
\nonumber &=& \gamma^3m_o\frac{d(\delta \texttt{v})}{dt} 
\end{eqnarray}

Therefore the inertial mass of the particle is again $\gamma^3m_o$ and thus eq. (\ref{eq:3}) is again valid, i.e.:
\begin{equation*}
m_g=m_i=\gamma^3m_o
\end{equation*}

Consequently the values of the particle rest mass, $m_o$, and of its velocity, $\texttt{v}$, are sufficient to determine the inertial mass of the particle. Since the laboratory observer $S$ examining the instantaneous frame $S'$  (which is defined solely by the velocity vector $\textbf{v}$) cannot distinguish if the particle under consideration is performing a linear or non-linear (e.g. cyclic) motion (Fig. 2), one may reasonably conclude that in both cases, (which are characterized by the same $\textbf{v}$, linear motion or nonlinear motion) the inertial mass is the same, i.e. $\gamma^3m_o$. 

It is worthwhile to confirm this by starting from the relativistic equation of motion (eq. \ref{eq:8}) and examining the general case where $\textbf{F}$ and $\textbf{v}$ are not parallel. 

\textbf{C. Arbitrary force}: We thus return to the general case where a non zero force, $\textbf{F}$, is already acting on the particle which is in general not parallel to $\textbf{v}$. 

We remind again that when $\textbf{F}$ and $\textbf{v}$ are not parallel, then the force, $\textbf{F}$, is not invariant, i.e. different values of it are perceived in $S$ and $S'$ and thus the actual force value acting on the particle is not the same with that perceived by the laboratory observer. 

Consequently the laboratory observer cannot utilize directly the relativistic equation of motion (\ref{eq:8}):
\begin{equation*}
\textbf{F}=\frac{d\textbf{p}}{dt}=\gamma m_o\frac{d\textbf{v}}{dt}+\gamma^3m_o\frac{1}{c^2}\left(\textbf{v}\bullet\frac{d\textbf{v}}{dt}\right)\textbf{v}
\end{equation*}
to compute the real inertial mass, since $F$ is not invariant, i.e. uniquely defined in $S$ and $S'$. To overcome this difficulty one can superimpose to the actual force $\textbf{F}$ an infinitesimal test force $\delta\textbf{F}$ colinear to the instantaneous velocity vector $\textbf{v}$ \cite{French68}, (Fig. 3) regardless of the actual motion (e.g. linear or circular) performed by the particle. The reason for choosing the test force $\delta\textbf{F}$ parallel to the instantaneous velocity $\textbf{v}$ is that, as already discussed, the case where the force, $\textbf{F}$, and $\textbf{v}$ are colinear is the only one for which the force is invariant, i.e. the only one where the same value of $\textbf{F}$ is percieved in $S$ and $S'$. It is obvious that if $\textbf{F}$ is not invariant, i.e. is not uniquely defined in $S$ and $S'$, any correlation between $\textbf{F}$ and $d\textbf{v}/dt$ is not meaningful.  The superposition of the infinitesimal test force $\delta\textbf{F}$ parallel to $\textbf{v}$ causes an infinitesimal change, $\delta\textbf{v}$, in the particle velocity, in the direction parallel to $\textbf{v}$. Since  $\textbf{v}$ and $\delta\textbf{v}$ are colinear it follows:
\begin{equation}
\left[\textbf{v}\bullet\frac{d\delta\textbf{v}}{dt}\right]\textbf{v}=\texttt{v}^2\frac{d(\delta\textbf{v})}{dt}
\label{eq:13}
\end{equation}
\begin{figure}[ht]
\begin{center}
\includegraphics[width=0.40\textwidth]{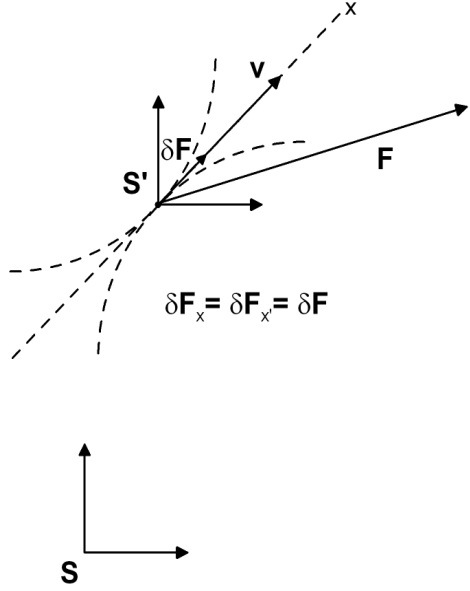}
\caption{Laboratory frame $S$ and instantaneous inertial frame $S'$, the latter moving with the particle under consideration. The frame $S'$ is uniquely defined by the vector $\textbf{v}$ alone regardless of the motion (e.g. linear or circular) performed by the particle. The test force $\delta\textbf{F}$ is applied parallel to the instantaneous velocity $\textbf{v}$;  $\delta\textbf{F}_x$ and  $\delta\textbf{F}_{x'}$ are the test force values perceived in the reference frames $S$ and $S'$ \cite{French68} denotes the direction of $\textbf{v}$. }
\label{3}
\end{center}
\end{figure}

Due to the superposition of this infinitesimal test force $\delta\textbf{F}$, equation (\ref{eq:8}) can now be written as:
\begin{equation}
\textbf{F}+\delta\textbf{F}=\gamma m_o\frac{d(\textbf{v}+\delta\textbf{v})}{dt}+\gamma^3m_o\frac{1}{c^2}\left[(\textbf{v}+\delta\textbf{v})\bullet\frac{d(\textbf{v}+\delta\textbf{v})}{dt}\right](\textbf{v}+\delta\textbf{v})
\label{eq:14}
\end{equation}

Subtracting equation (\ref{eq:8}) from (\ref{eq:14}), negleting $(\delta\textbf{v})^2$ terms and using equation (\ref{eq:13}) one obtains:
\begin{equation}
\delta\textbf{F}=\gamma m_o\frac{d(\delta\textbf{v})}{dt}+\gamma^3m_o\frac{\texttt{v}^2}{c^2}\frac{d(\delta\textbf{v})}{dt} 
\label{eq:15}
\end{equation}
which, since $\delta\textbf{F}$ and $\delta\textbf{v}$ are colinear, reduces to:
\begin{equation}
\delta F=m_o\left[\gamma + \gamma^3\frac{\texttt{v}^2}{c^2}\right]\frac{d(\delta\textbf{v})}{dt} 
\label{eq:16}
\end{equation}

Accounting again for $\gamma=(1-\texttt{v}^2/c^2)^{-1/2}$ this equation reduces to: 
\begin{equation}
\delta F=\gamma^3m_o\frac{d(\delta\textbf{v})}{dt} 
\label{eq:17}
\end{equation}

Note that the non-parallel to $\textbf{v}$ and thus not invariant force $\textbf{F}$ has been eliminated and therefore, as in the case of the linear motion, one obtains: 
\begin{equation}
m_i=\delta F/(d(\delta\textbf{v})/dt) =\gamma^3m_o
\label{eq:18}
\end{equation}

Consequently the use of instantaneous reference frames in conjunction with the relativistic equation of motion leads to the same expression for the inertial mass for arbitrary particle motion, as that already derived for linear motion. 

In view of equation (\ref{eq:18}) one may now proceed to compute the gravitational force via Newton's universal gravitational law by using the actual gravitational masses of the two attracting particles. Thus upon considering a second particle of rest mass $m_o$ and instantaneous speed v relative to the observer at $S$ and at a distance $r$ from the first particle, it follows that the inertial and gravitational mass of the second particle is also given by $\gamma^3m_o$, as in equation (\ref{eq:18}), and thus one can use these two $m_g$ values in Newton's universal gravitational law in order to compute the gravitational force, $F_G$, between the two particles. Thus from:
\begin{equation} 
\label{eq:19} 
F_{G}=-\frac{Gm_{1,g}m_{2,g}}{r^{2}}
\end{equation}
and equation(\ref{eq:18}) one obtains:
\begin{equation} 
\label{eq:20} 
F_{G}=-\frac{Gm_{o}^{2} \gamma ^{6}}{r^{2}}
\end{equation}
which depends on the $6^{th}$ power of $\gamma$ and accounts explicitly for the velocity dependence of the inertial and gravitational mass. It is worth remembering that this equation stems directly from special relativity (eq. \ref{eq:2} or \ref{eq:8}), the weak equivalence principle (eq. \ref{eq:3} or \ref{eq:10}) and Newton's gravitational law. No other assumptions are involved except that Newton's universal gravitational law remains valid under relativistic conditions via the use of the relativistic inertial and thus gravitational mass. 

Application of equation (\ref{eq:20}) to the circular motion of Figure 1 gives after some simple trigonometry:
\begin{equation}
\label{eq:21}
F_{G}=-\frac{Gm_{o}^{2}\gamma ^{6}(R)}{\sqrt{3} {\rm \;}R^{2}}
\end{equation}

\subsection{The classical mechanical problem}
Upon combining with equation (\ref{eq:7}) one obtains:
\begin{equation} 
\label{eq:22} 
\frac{\gamma (R)m_{o}\texttt{v}^{2}}{R}=\frac{Gm_{o}^{2}\gamma^{6}(R)}{\sqrt{3}{\rm \; }R^{2}}
\end{equation}
i.e., the gravitational force $F_G$ given by equation (\ref{eq:21}) acts as the centripetal force for the rotational motion. 

It must be noted that on the basis of (\ref{eq:8}) one might be tempted to assign the value $\gamma m_o$, commonly termed transverse mass \cite{French68,Freund08}, to the inertial and thus gravitational mass of each particle. However, as already noted, the mass $m_i(=m_o\gamma^3)$, and thus also $m_g$, is uniquely determined for given $m_o$ and $\textbf{v}$, via the colinear to $\textbf{v}$ test force $\delta\textbf{F}$, and does not depend on the type of motion (e.g. linear or circular) performed by the particle. One notes that in general the relativistic mass, $\gamma m_o$, is not the same with the inertial and thus gravitational mass, $\gamma^3m_o$.

Upon utilizing $\gamma (R)=(1-\texttt{v}^{2} /c^{2} )^{-1/2}$ in eq. (\ref{eq:22}) one obtains:
\begin{equation} 
\label{eq:23} 
R=\frac{Gm_{o}}{\sqrt{3}{\rm \; c}^{{\rm 2}} } \gamma ^{5} \left(\frac{\gamma ^{2}}{\gamma^{2}-1}\right)
\end{equation}
or, equivalently:
\begin{equation} 
\label{eq:24} 
R=(R_{S}/(2\sqrt{3}))\gamma ^{5} \left(\frac{\gamma ^{2}}{\gamma ^{2}-1} \right)
\end{equation}
where $R_{S}(=2Gm_{o}/c^{2})$ is the Schwarzschild radius of a particle with rest mass $m_o$.
\begin{figure}[ht]
\begin{center}
\includegraphics[width=0.45\textwidth]{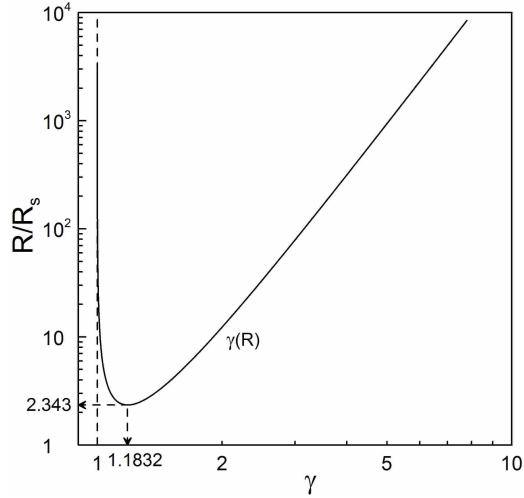}
\end{center}
\caption{Plot of eq. (\ref{eq:24}) near the minimum R.}
\label{fig:4}
\end{figure}

As shown in Figure 4 the function R defined by eq. \ref{eq:24} exhibits a minimum, $R_{\min}=2.343R_{S} $, at $\gamma _{\min}=\sqrt{7/5} =1.1832$ thus v$_{\min}=\sqrt{2/7}c$. This is the minimum radius for a circular orbit and the corresponding minimum angular momentum is $L_{\min}=\gamma _{\min}m_{o}\texttt{v}_{\min}R_{\min}=1.481m_{o}cR_{S}=2.963$ $Gm_{o}^{2}/c$.

The above condition, i.e. $L>2.963\;Gm^2_o/c$, is similar to the criteria $L>Gm^2/c$ found for circular orbits in special relativity \cite{Torkelsson98,Boyer04} or $L>2\sqrt{3} Gm^{2} /c$ for the Schwarzschild metric in general relativity \cite{Boyer04} with orbits around point masses with $r^{-1}$ potentials.

Equation (\ref{eq:24}) defines two $\gamma $ branches (Fig. 4), one corresponding to low $\gamma$ values $(\gamma <1.1832)$ the other corresponding to large $\gamma$ values $(\gamma > 1.1832)$. The first branch corresponds to common Keplerian gravitational orbits. In this case $\gamma$ and thus the velocity $\textbf{v}$ decreases with increasing R, e.g. $\texttt{v}=(Gm_{o}/(\sqrt{3}R))^{1/2}$ in the non-relativistic case $(\gamma \approx 1)$.

The second branch which leads to relativistic velocities defines rotational states where $\gamma$ and thus v increase with increasing R. These states with $\gamma \gg 1$ are the states of primary interest here, since they correspond to high velocities and strong gravitational attraction. For $\gamma \gg 1$, e.g. $\gamma >10^{2}, $ eq. (\ref{eq:24}) reduces to:
\begin{equation} 
\label{eq:25}
R=(R_{S}/(2\sqrt{3}))\gamma^{5}\quad;\quad \gamma=(2\sqrt{3})^{1/5}(R/R_S)^{1/5}
\end{equation}

\subsection{The neutrino de Broglie wavelength and the mass of the bound state}
In analogy to the Bohr model of the H atom one can then proceed to identify among the infinity of bound rotational states described by eq.(\ref{eq:25}), each corresponding to a different $R$, those rotational states where $R$ is an odd integer multiple of the reduced de Broglie wavelength  ${\mathchar'26\mkern-10mu\lambda}(=\hbar /p)$ of the light rotating particles.

The latter is computed from:
\begin{equation} 
\label{eq:26}
\mathchar'26\mkern-10mu\lambda=\frac{\hbar}{p}=\frac{\hbar}{\gamma m_o\texttt{v}}\approx\frac{\hbar}{\gamma m_oc}
\end{equation}

Thus assuming $R=(2n-1)\mathchar'26\mkern-10mu\lambda$ one obtains:
\begin{equation} 
\label{eq:27}
R=\frac{(2n-1)\hbar}{\gamma m_oc}=\frac{3(2n-1)\hbar}{mc}
\end{equation}
where $m(=3\gamma m_o)$ is the mass of the bound rotational state formed.

Similarly to the Bohr model of the H atom, this is mathematically equivalent to introducing quantization of the angular momentum or of the action of the neutrinos in the form:
\begin{equation} 
\label{eq:28}
L=\gamma m_{o}Rc=(2n-1)\hbar
\end{equation}
which is identical to eq. (\ref{eq:27}). Using (\ref{eq:27}) and the definition of $R_S(=2Gm_o/c^2)$ one can thus express the ratio $R/R_S$ in the form:
\begin{equation} 
\label{eq:29}
\frac{R}{R_S}=\frac{(2n-1)\hbar c}{2\gamma Gm^2_o}
\end{equation}

This ratio is also given by equation (\ref{eq:25}), i.e. 
\begin{equation} 
\label{eq:30}
\frac{R}{R_S}=\frac{\gamma^5}{2\sqrt{3}}
\end{equation}

Consequently from (\ref{eq:29}) and (\ref{eq:30}) one obtains:
\begin{equation} 
\label{eq:31}
\gamma^6=\frac{3^{1/2}(2n-1)\hbar c}{Gm^2_o}=3^{1/2}(2n-1)\frac{m^2_{Pl}}{m^2_o}
\end{equation}
where $m_{Pl}=(\hbar c/G)^{1/2}$ is the Planck mass. Recalling that the mass, $m$, of the bound rotational state equals $3\gamma m_o$ one thus obtains:
\begin{equation} 
\label{eq:32}
m=3\gamma m_o=3^{13/12}(2n-1)^{1/6}m^{2/3}_om^{1/3}_{Pl}
\end{equation}
i.e. the mass of the bound state has been expressed in terms of $m_o$ and natural constants. This completes the solution of the three rotating neutrino model.

\subsection{Numerical substitutions}
Upon setting $n=1$ and $m_o=0.04\;eV/c^2$ in equation (\ref{eq:32}) and using the Planck mass value $m_{Pl}=1.221\cdot 10^{19}\;GeV/c^2$ one obtains:
\begin{equation} 
\label{eq:33}
m=885.43\; MeV/c^2
\end{equation}
which, surprisingly, is in the baryons mass range and in fact differs less than 6\% from the rest mass of the proton (938.272 $MeV/c^2$) and of the neutron (939.565 $MeV/c^2$). Exact agreement with the neutron mass, $m_n$, is obtained for: 
\begin{equation} 
\label{eq:34}
m_o=0.043723\; eV/c^2=7.7943\cdot 10^{-38} \; kg
\end{equation}
This is the value computed from equation (\ref{eq:32}) for $n=1$ which is assumed to correspond to a neutron, i.e.:
\begin{equation} 
\label{eq:35}
m_o=\frac{(m_n/3)^{3/2}}{3^{1/8}m^{1/2}_{Pl}}
\end{equation}
The thus computed $m_o$ value is in quite good agreement with the current best estimate of $m_o=0.051(\pm0.01)\;eV/c^2$ for the mass of the heaviest neutrino \cite{Mohapatra07} extracted from the Super-Kamiokande data \cite{Mohapatra07}. This value is computed from the square root of the $\left|\Delta m^2_{23}\right|$ value of $2.6(\pm0.2)\times 10^{-3}(eV/c^2)^2$ extracted from the Super-Kamiokande data for the $\nu_\mu\:\longleftrightarrow\:\nu_\tau$ oscillations \cite{Mohapatra07}.

Actually as shown in Figure 5 the $m_o$ value of 0.043723 $eV/c^2$ (eqs. \ref{eq:34} and \ref{eq:35}) practically coincides with the currently computed maximum neutrino mass value both for the normal mass hierarchy ($m_3>>m_2>m_1$) and for the inverted hierarchy ($m_1\approx m_2>>m_3$) \cite{Mohapatra07}.
\begin{figure}[ht]
\begin{center}
\includegraphics[width=0.90\textwidth]{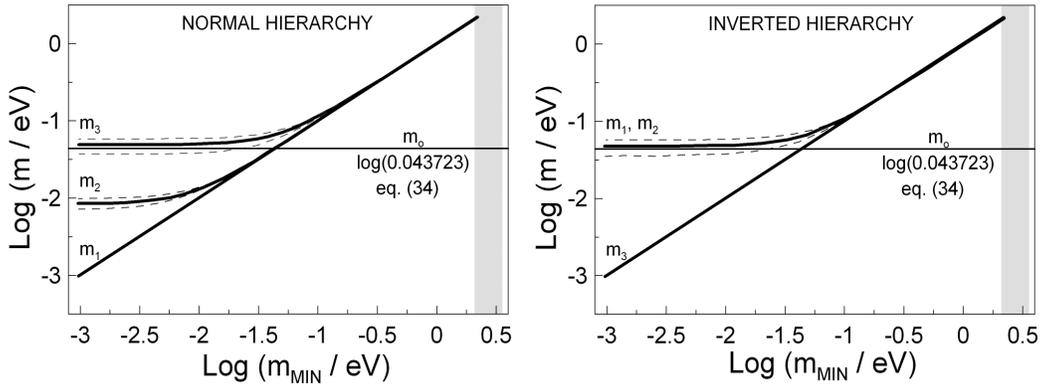}
\caption{The three light neutrino masses as a function of the lightest mass for the normal (top plot) and inverted (bottom plot) hierarchy, reprinted from ref. \cite{Mohapatra07} and comparison with equation (\ref{eq:34}) or (\ref{eq:35}), i.e. $m_o=(m_n/3)^{3/2}/(3^{1/8}m^{1/2}_{Pl})=0.043723\;eV/c^2$.}
\label{fig:5}
\end{center}
\end{figure}

\begin{figure}[ht]
\begin{center}
\includegraphics[width=8.0cm,height=7.0cm]{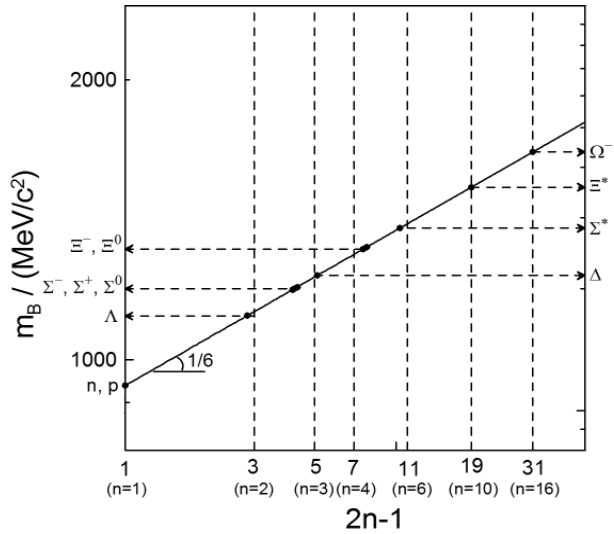}
\caption{Comparison of the masses, $m_B$, of the uncharmed baryons, consisting of u, d and s quarks, \cite{Griffiths08} with equation \ref{eq:36}, i.e. $m_B=m_n(2n-1)^{1/6}$ where $m_n$ is the neutron
mass.} \label{fig:6}
\end{center}
\end{figure}

It is worth noting that for any fixed $m_o$ value, equation (\ref{eq:32}) can also be written as: 
\begin{equation} 
\label{eq:36}
m=(2n-1)^{1/6}m_n
\end{equation}
where $m_n$ is the neutron mass. As shown in Figure 6 this expression is also in very good agreement with experiment regarding the masses of baryons consisting of $u$, $d$ and $s$ quarks \cite{Griffiths08} which follow the $(2n-1)^{1/6}$ dependence of equation (\ref{eq:36}) with an accuracy better than 3\% (Fig. 6 and Table 1).
\begin{table}
\begin{center}
\caption{Experimental \cite{Griffiths08} and computed (eq. \ref{eq:36}) baryon masses}
\begin{tabular}{|p{0.60in}|p{1.5in}|p{1.00in}|p{0.3in}|}\hline
Baryon & Experimental mass value $MeV/c^2$ & $m_n(2n-1)^{1/6}$ & n \\ \hline
\vspace{0.1cm}$N \left\{\begin{array}{l}{p}\\{n}\end{array}\right.$ & \vspace{0.1cm} 938.272\newline 939.565 & \vspace{0.1cm}939.565 &\vspace{0.1cm} 1 \\ \hline 
$\Lambda$ & 1115.68 & 1128.3 & 2 \\ \hline
\vspace{0.1cm}$\Sigma^{+}$\newline $\Sigma^{o}$\newline $\Sigma^{-} $\newline $\Delta$ &\vspace{0.1cm} $\left.\begin{array}{l}1189.37\\1192.64\\1197.45\\1232\end{array}\right\}$\vspace{0.1cm} &\vspace{0.8cm}1228.6 & \vspace{0.6cm}3\\ \hline 
$\Xi ^{o} $\newline $\Xi ^{-} $ & 1314.8\newline 1321.3 & 1299.5 & 4\\ \hline
$\Sigma *$ & 1385 & 1401.2 & 6 \\ \hline
$\Xi *$ & 1533 & 1534.7 & 10 \\ \hline
$\Omega ^{-}$ & 1672 & 1665.3 & 16 \\ \hline
\end{tabular}
\end{center}
\end{table}
As already noted (e.g. equation \ref{eq:32}) the total rest plus kinetic energy of the three rotating particles equals $3\gamma m_oc^2$ and constitutes at the same time the rest energy, $mc^2$, of the composite particle formed, i.e. of the rotational bound state:
\begin{equation} 
\label{eq:37} 
mc^2=3\gamma m_oc^2
\end{equation}

It is useful to note that in the model the rest mass of the three individual particles, i.e. $3m_oc^2$, does not change when the bound state is formed. The transformation of the kinetic energy of the three rotating particles into rest energy of the bound state is associated with the change in choice of the boundaries of the system. In the former case (three individual rotating particles) the boundaries are geometrically disconnected and encompass each particle individually, in the latter case the system boundary contains all three particles and the center of mass of the confined state is at rest with respect to the laboratory observer. Thus the formation of the bound rotating state by the three relativistic particles provides a simple hadronization mechanism, i.e. a mechanism for generation of rest mass, $m$, starting from an initial rest mass $3m_o$, according to eq. (\ref{eq:37}).

\section{Other properties of the bound rotational states}
The Bohr-de Broglie type rotational model presented in section 2 has shown that the bound rotational state formed by three gravitationally attracting neutrinos of rest mass $m_o=0.043723\; eV/c^2$ each, has a rest mass of 939.565 $MeV/c^2$, equal to that of the neutron. This is at first quite surprising and one might think that it could be fortuitous. 

It is therefore useful to examine some other key properties of the rotational neutrino states and compare with experiment. 

\subsection{Potential, translational and total energy}
In order to compute the binding energy of the bound state it is necessary to return to the force expression eq. (\ref{eq:21}) and to use eq. (\ref{eq:25}) in order to eliminate $\gamma $ in eq. (\ref{eq:21}). One thus obtains:
\begin{equation} 
\label{eq:38} 
F_{G} =-m_{o} c^{2} \left(\frac{2\sqrt{3} }{R_{S} } \right)^{1/5} \frac{1}{R^{4/5} }
\end{equation}

The force equation (\ref{eq:38}) refers to circular orbits only and thus defines a certain conservative force, since the work done in moving the particles between two points $R_1$ and $R_2$, corresponding to two rotational states with radii $R_1$ and $R_2$, is independent of the path taken. The force vector orientation is also given, pointing to the center of rotation, and thus a conservative vector field is defined which is the gradient of a scalar potential, denoted $V_G(R)$. The latter is the gravitational potential energy of the three rotating particles when accounting for their rotational motion and corresponds to the energy associated with transfer of the particles from the minimum circular orbit radius $R_{\min }$ to an orbit of radius of interest, $R$. The function $V_G(R)$ is obtained via integration of eq. (\ref{eq:38}), i.e.,
\begin{eqnarray}
\label{eq:39}
V_{G}(R)-V_{G}(R_{min})&=&-\int_{R_{min}}^{R}F_{G}dR'=\\&=&
\nonumber -5m_{o}c^{2} \left(\frac{2\sqrt{3}}{R_{S}}\right)^{1/5}\left(R^{1/5}-R_{\min}^{1/5}\right)
\end{eqnarray}

Noting that $R_{min}=2.343R_S$ (Fig. 4) and that the value of the Schwarzchild radius, $R_S$, $(=2Gm_o/c^2)$ for neutrinos is extremely small $(\sim 10^{-63}\:m)$ it follows that for any realistic $R$ value (e.g. above the Planck length value of $10^{-35}\:m)$ equation (\ref{eq:39}) reduces to:
\begin{equation} 
\label{eq:40}
V_{G}(R)=-5m_{o}c^{2}(2\sqrt{3})^{1/5}(R/R_{S})^{1/5}
\end{equation}

Thus while the magnitude of the gravitational force acting on the rotating particles increases with decreasing radius, R (eq. \ref{eq:38}), the absolute value $\left|V_G(R)\right|$ of the gravitational potential energy increases constantly with increasing $R$ and is unbound (eq. \ref{eq:40}). Therefore equation (\ref{eq:40}) describes confinement which is one of the main characteristics of the strong force \cite{Hooft07,Gross73,Politzer73,Cabibbo75}. The same equation (\ref{eq:40}) also describes asymptotic freedom \cite{Gross73,Politzer73,Cabibbo75}, i.e. the attractive interaction energy becomes very small at short distances, which is a second key characteristic of the strong force \cite{Gross73,Politzer73,Cabibbo75}. 

In view of Eq. (\ref{eq:25}) one can rewrite equation (\ref{eq:40}) as:
\begin{equation} 
\label{eq:41}
V_{G}(R)=-5\gamma m_{o} c^{2}
\end{equation}

On the other hand the total kinetic energy, $T$, of the three rotating neutrinos is:
\begin{equation} 
\label{eq:42}
T(R)=3(\gamma-1)m_{o}c^{2}
\end{equation}

Thus one may now compute the change, $\Delta \mathcal{H}$, in the Hamiltonian \index{Hamiltonian}, $\mathcal{H}$, i.e. in the total energy \index{total energy} of the system, upon formation of the rotational bound state \index{rotational bound state} from the three originally free neutrinos. The Hamiltonian, $\mathcal{H}$, is the sum of the relativistic energy \index{relativistic energy}, $E=3 \gamma m_oc^2$, and of the potential energy \index{potential energy} $V_G$. The relativistic energy is the sum of the rest energy, $3m_oc^2$, and of the kinetic energy $T$. Denoting by f and i the final and initial states (i.e. the three free non-interacting neutrinos at rest and the bound rotational state) and by (RE) the rest energy, one obtains:
\begin{eqnarray} 
\label{eq:43}
\Delta \mathcal{H}&=&\mathcal{H}_{f}-\mathcal{H}_{i}=\\
\nonumber &=&\left[(RE)_{f}+T_{f}+V_{G,f}\right]-\left[(RE)_{i}+T_{i}+V_{G,i} \right]=\\
\nonumber &=&\left[3m_{o}c^{2}+3(\gamma -1)m_{o}c^{2}-5\gamma m_{o}c^{2}\right]-3m_{o} c^{2}=\\ 
\nonumber &=&\Delta T+\Delta V_{G} =-(2\gamma +3)m_{o} c^{2} \approx -2\gamma m_{o} c^{2}
\end{eqnarray}
where the last equality holds for $\gamma \gg 1$ as is the case of interest here.

The negative sign of $\Delta \mathcal{H}$ shows that the formation of the bound rotational state starting from the three initially free neutrinos happens spontaneously, is exoergic $(\Delta \mathcal{H}<0)$, and the binding energy $BE(=-\Delta \mathcal{H})$ equals $2\gamma m_oc^2$.

It follows from (\ref{eq:37}), i.e. $mc^2=3\gamma m_oc^2$, and from (\ref{eq:43}), that the binding energy, $BE$, can be computed from:
\begin{equation} 
\label{eq:44}
BE=-\Delta \mathcal{H}=(2/3)mc^{2}
\end{equation}

Thus the binding energy per light particle is $(2/9)mc^2$, which for $m=m_p=938.272\;MeV/c^2$, the proton mass, gives an energy of 208 $MeV$, in good qualitative agreement with the estimated particle energy of 150-200 $MeV$ at the transition temperature of QCD \cite{Braun07} and in even better agreement with the QCD scale of $217 \pm 25\;MeV$   \cite{Wilczek04}.

One may note here that the potential energy expression (\ref{eq:41}) can be shown easily not to depend on the number, $N$, of rotating particles. On the other hand the kinetic energy, T, is a linear function of $N$, i.e. $N(\gamma -1)m_oc^2$. Thus it follows from (\ref{eq:43}) that stable rotational states $(\Delta \mathcal{H}<0)$ cannot be obtained for $N>5$ since they lead to positive $\Delta \mathcal{H}$. The case $N=2$ is also interesting, as is the case $N=3$ treated here. It can be shown, in a way similar to that presented in sections 2.2 and 2.3, that the case $N=2$ leads to composite masses, $m$, in the range of mesons, i.e. in the $0.5\;GeV/c^2$ range \cite{Vayenas12}.

\subsection{Energy-curvature dependence and general relativity}
Since the present Bohr-type model is based on the combination of special relativity and the equivalence principle, which was the basis of the theory of general relativity, it is worthwhile to explore if the simple mathematical equations of the present model may have some similarity with some limiting form of the field equations of general relativity.

Thus by using equation (\ref{eq:25}) to eliminate $R$ in the force expression of equation (\ref{eq:21}) one obtains:
\begin{equation} 
\label{eq:45}
F_G=-\sqrt{3}\frac{c^4}{\gamma^4 G}
\end{equation}
and thus, interestingly, for any given value of $\gamma$, and thus, via eq. (\ref{eq:25}), for any given value of $R$, the attractive force is uniquely determined by the familiar $G/c^4$ parameter of the gravitational field equations of general relativity, i.e.:
\begin{equation} 
\label{eq:46}
G_{ik}=8\pi\left(\frac{G}{c^4}\right)T_{ik}
\end{equation}
which relates the Einstein tensor $G_{ik}$ with the stress-momentum-energy tensor $T_{ik}$ \cite{Wheeler55,Misner73,Misner09,Hawking06}.

In view of equation (\ref{eq:39}) the force, $F_G$, can also be expressed as:
\begin{equation} 
\label{eq:47}
F_G=-\frac{dV_G}{dR}
\end{equation}
and thus combining with (\ref{eq:45}) one obtains:
\begin{equation} 
\label{eq:48}
dR=\frac{\gamma^4}{\sqrt{3}}\left(\frac{G}{c^4}\right)dV_G
\end{equation}
which for small variations in $\gamma$ and thus $R$ and $V_G$ gives:
\begin{equation} 
\label{eq:49}
\Delta R=\frac{\gamma^4}{\sqrt{3}}\left(\frac{G}{c^4}\right)\Delta V_G
\end{equation}

Upon comparing with the field equations (\ref{eq:46}) one observes that (\ref{eq:49}) is similar to a limiting one-dimensional analogue of (\ref{eq:46}) with the change in radius, or in curvature $\Delta R$, being analogous to the spacetime curvature due to the presence of mass and the change in gravitational energy, $\Delta V_G$, being analogous to the stress-momentum-energy tensor $T_{ik}$. 

\subsection{Radii and Lorentz factors \textbf{$\gamma$}}
The rotational radius, $R$, of the bound state computed from equation (\ref{eq:27}) for $n=1$, i.e.
\begin{equation} 
\label{eq:50}
R(n=1)=\frac{3 \hbar}{m_nc}=0.631\;fm
\end{equation}
is the neutrino de Broglie wavelength in the bound state and equals three times the neutron Compton wavelength. This value is in very good agreement with the experimental proton and neutron radii values which lie in the 0.6 - 0.7 $fm$ range. 

For $n>1$ the corresponding $R(n)$ values can be computed from equation (\ref{eq:27}), i.e. 
\begin{equation} 
\label{eq:51}
R=\frac{(2n-1)\hbar}{\gamma m_oc}
\end{equation}
By accounting for the $\gamma$ dependence on $(2n-1)$ given by equation (\ref{eq:31}), i.e.
\begin{equation} 
\label{eq:52}
\gamma(n)=(2n-1)^{1/6}\gamma(n=1)=7.169\cdot 10^9(2n-1)^{1/6}
\end{equation}
one obtains:
\begin{equation} 
\label{eq:53}
R(n)=(2n-1)^{5/6}R(n=1)=0.631(2n-1)^{5/6}fm
\end{equation}

The $\gamma(n)$ values are in the range computed in the Introduction for 200 $MeV$ neutrinos. The radii $R(n)$ also lie in the range of hadron, e.g. proton or neutron, radii.

\subsection{Lifetimes and rotational periods}
The period of rotation $\tau(n)$ of the neutrinos within the composite state, $2\pi R/\texttt{v}\sim 2\pi R/c$, is, using Eq. \ref{eq:53},
\begin{equation} 
\label{eq:54}
\tau (n)=(2n-1)^{5/6} \tau _{p}=(2n-1)^{5/6} 6.6\cdot 10^{-24} \;s
\end{equation}
where $\tau_{p}=2\pi R_{p}/c=6.6\times 10^{-24} \;s$ is the rotation period for the proton or the neutron.  The time interval $\tau(n)$ provides a rough lower limit for the lifetime of the composite particles, interpreted as baryons, as they can be defined only if the neutrinos complete at least a revolution (Fig. 1). Indeed all the known lifetimes of the baryons are not much shorter than that estimate. The lifetime of the $\Delta $ baryons, which is the shortest, is $5.6\cdot 10^{-24} s$ \cite{Griffiths08}.

\subsection{Spins and charges}

Neutrinos are fermions with spin $1/2$ \cite{Griffiths08} and thus one may anticipate spin of $1/2$, or $3/2$ for composite states formed by three neutrinos. Indeed most baryons have spin $1/2$ and some have spin $3/2$ \cite{Griffiths08}. If the bound state discussed in the present model involves two neutrinos and one antineutrino, then a spin of $1/2$, that of a neutron, can be anticipated for the bound state.  

Several baryons are charged, e.g. the proton or the $\Xi^+$. The differences in mass, $m$, from their neutral brethren (i.e. the neutron or the $\Xi^o$) are small and of the order $\alpha m$, where $\alpha(=e^2/\epsilon c\hbar=1/137.0359)$ is the fine structure constant. Thus the rotating neutrino model discussed here can describe with reasonable accuracy (e.g. Fig. 6) the masses of both neutral and charged baryons. However, since neutrinos are electrically neutral, the question arises about how charged baryons can be formed within the rotating neutrino model. 

It appears possible that electrically neutral baryons (e.g. neutrons) can be first formed from neutrinos and that these neutral baryons can subsequently be transformed to charged baryons (e.g. protons) via the $\beta$-decay:
\begin{equation} 
\label{eq:55}
n\rightarrow p^++e^-+\overline{\nu}_e
\end{equation}
where $\overline{\nu}_e$ is the electron antineutrino. The half life of this reaction is 885.7 $s$ \cite{Griffiths08}. Other possibilities may also be sought. Regarding the distribution of charge in the composite state one may assume that the charges of the constituent particles are equal to those of $u$ and $d$ quarks, i.e. $(2/3)e$ and $-(1/3)e$. This leads as shown in the next section to very good agreement with experiment regarding magnetic moments. 

With this assumption one can estimate the Coulomb interaction energy between the rotating particles. In the simplified geometry of Figure 1 the total Coulomb potential energy for the proton (charges 2/3, 2/3, -1/3) vanishes, i.e. denoting $\varepsilon =4\pi \varepsilon_{o}$ one obtains: 
\begin{equation} 
\label{eq:56}
V_{C,p}=\frac{e^{2}}{\varepsilon \sqrt{3}R}\left[(4/9)-(2/9)-(2/9)\right]=0
\end{equation}
while for the neutron (charges -1/3, -1/3, 2/3) it is negative: 
\begin{equation} 
\label{eq:57}
V_{C,n}=\frac{e^{2}}{\varepsilon \sqrt{3}R}\left[(1/9)-(2/9)-(2/9)\right]=-\frac{(e^{2} /\varepsilon)}{3\sqrt{3}R}
\end{equation}
, i.e. there is an overall attractive Coulombic interaction.

Upon substituting $R$ from Eq. (\ref{eq:27}) for the case of the neutron $(n=1)$ one obtains:
\begin{equation} 
\label{eq:58}
V_{C,n}=-\frac{e^{2} }{9\sqrt{3} \varepsilon c\hbar} m_{p} c^{2}=-\frac{\alpha}{9\sqrt{3}}m_{n}c^{2}=-4.69\cdot 10^{-4}\;m_nc^2=-0.44\; MeV/c^2
\end{equation}
which confirms that the Coulombic interaction energy is negligible in comparison to the relativistic gravitational interaction energy and is of the same order of magnitude as the difference in the rest energies ($\sim 1.3\;MeV/c^2)$ of neutrons and protons. 

Nevertheless if the Coulomb interaction is taken into consideration the symmetry of the configuration of Fig. 1 is broken as not all three charges are the same. Although the deviation from three-fold symmetry is small, since the Coulombic energy is small, and thus one may still use with good accuracy eq. (\ref{eq:57}) to estimate the attractive interaction between the three particles forming a neutron, it is conceivable that this broken symmetry may be related to the relative instability of the neutron (lifetime 885.7 $s$) vs the proton (estimated lifetime $\sim 10^{32}\;s$ \cite{Griffiths08}).

\subsection{Magnetic moments}
It is interesting to compute the magnetic dipole moments, $\mu$, of the bound rotational states. Using the definition of $\mu (=(1/2)qR$v) and considering the case $n=1$, corresponding to a proton (which is a uud baryon), with charge $2e/3$ for u and $-e/3$ for d it follows:
\begin{equation} 
\label{eq:59}
\mu_{p}=(1/2)eRc\left[(2/3)+(2/3)-(1/3)\right]=(1/2)eRc
\end{equation}

Upon substituting $R=R_{p}=0.631 \; fm$ one obtains:
\begin{equation}
\label{eq:60}
\mu_{p}=15.14\cdot 10^{-27} {\rm \; J/T\; \; \; \; \; (=3\mu_N)}
\end{equation}
where $\mu _{N} $ is the nuclear magneton $(5.05\cdot 10^{-27} {\rm \; }J/T)$. This value differs less than 8\% from the experimental value of $14.10\cdot 10^{-27} {\rm \; J/T\; \; (i.e.\; 2.79\; }\mu _{{\rm N}} )$ \cite{Mohr05}.

In the above computation (eq. \ref{eq:59}) one assumes that the spin vectors of the three small particles (i.e. uud) are parallel to the vector of rotation of the rotating proton state. If one considers the neutron which is a udd particle and assumes that the spin of one of the two d quarks is parallel with the rotation vector of the rotating neutron state and the spins of the other two particles are antiparallel to the neutron rotation vector then one obtains:
\begin{equation}
\label{eq:61}
\mu_{n}=(1/2)eRc\left[(-2/3)+(1/3)-(1/3)\right]=-(1/3)eRc
\end{equation}
and upon substitution of $R=0.631\; fm$ one obtains:
\begin{equation} 
\label{eq:62}
\mu_{n}=-10.09\cdot 10^{-27}{\rm \; J/T}=-2\mu_{N}
\end{equation}
which is in excellent agreement with the experimental value of $-9.66\cdot 10^{-27} {\rm \; J/T\; }(=-1.913\mu _{N} )$.

This good agreement seems to imply that the spin contribution of the light particles to the magnetic moment of the rotating state is small and only the spin vector orientation (parallel or antiparallel to the baryon rotation vector) is important.

\subsection{Inertial mass and angular momentum}
Interestingly it follows from equation \ref{eq:31} that in the case of the neutron or proton $(n=1)$ the inertial and gravitational mass of each rotating particle, $\gamma^3m_o$, is related to the Planck mass, $m_{Pl}=(\hbar c/G)^{1/2}$, via a very simple equation, i.e. 
\begin{equation}
\label{eq:63}
\gamma^3m_o=3^{1/4}m_{Pl}=3^{1/4}\left(\frac{\hbar c}{G}\right)^{1/2}=1.607\cdot 10^{19}\quad GeV/c^2
\end{equation}
which provides an interesting direct connection between the Planck mass and the gravitational mass of the rotating neutrino model. The scale of gravity is generally expected to reach that of the strong force at energies approaching the Planck scale $(\sim 10^{19}\;GeV)$ \cite{Schwarz04} which is in good agreement with the model results (eq. \ref{eq:63}).

Thus while the relativistic mass, $3\gamma m_o$, of the bound state formed by the three neutrinos corresponds to $\sim 939\;MeV/c^2$, the inertial and gravitational mass  $\gamma^3 m_o$ of each of them is in the Planck mass range, i.e. $\sim 10^{19}\;GeV/c^2$ (Table 2). 
\begin{table}[hb]
\centering
\caption{Rest, relativistic, inertial and gravitational mass of the rotating neutrinos for $n=1$.}
\label{tab:2}     
\begin{tabular}{p{5.4cm}p{1.8cm}p{3.3cm}p{3.4cm}}
\hline\noalign{\smallskip}
 & Symbol & Value\\
\hline\noalign{\smallskip}
Rest mass & $m_{o}$  & $0.043723\;eV/c^2$\\
Relativistic mass & $\gamma m_{o}$ & $313.188\;MeV/c^2$ & (quark mass range)\\
Inertial mass or gravitational mass & $\gamma^3m_o$ & $1.60692\cdot 10^{19}\;GeV/c^2$ & (Planck mass range)\\ 
Confined state baryon mass $(n=1)$ &  $m=3\gamma m_o$ &  $939.565\;MeV/c^2$ & (neutron mass)\\
\hline\noalign{\smallskip}
\end{tabular}
\end{table}

It is worth reminding here Wheeler's concept of geons \cite{Wheeler55,Misner73,Misner09}, i.e. of electromagnetic waves or neutrinos held together gravitationally, which had been proposed as a classical relativistic model for hadrons \cite{Wheeler55}. Similarly to the present case (eq. (\ref{eq:63})) the minimum mass of a small geon formed from neutrinos had been estimated \cite{Wheeler55} to lie in the Planck mass range. 

It is interesting to note here that when using the inertial or gravitational mass, $\gamma^3m_o$, in the definition of the Compton wavelength, $\lambda_c$ of the particle $(=h/mc)$ then one obtains the Planck length $(\sim 10^{-35}\;m)$, but when using the mass corresponding to the total energy of the particles, $3\gamma m_o$, then one obtains the proton or neutron Compton wavelength $(\sim 10^{-15}\;m)$, which is close to the actual distance $(\sim fm)$ between the rotating particles.

The model is also qualitatively consistent with another central experimental observation about the strong force \cite{Hooft07}, i.e. that the normalized angular momentum of practically all hadrons and their excited states is roughly bounded by the square of their mass measured in $GeV$ \cite{Hooft07}. Indeed from eq. (\ref{eq:28}) and (\ref{eq:36}) one obtains:
\begin{equation}
\label{eq:64}
(L/\hbar)/(m/GeV)^2=1.13(2n-1)^{2/3}
\end{equation}
which is in reasonable qualitative agreement with experiment for small integer $n$ values.

\section{Conclusions}
A deterministic Bohr type model can be formulated for the rotational motion of three fast neutrinos using gravity as the attractive force. When accounting for special relativity, for the weak equivalence principle, for Newton's gravitational law and for the de Broglie wavelength of the rotating neutrinos, which leads to quantization of angular momentum, one finds that the rotational states formed have, surprisingly, the properties of baryons, including masses, radii, reduced Compton wavelengths, magnetic moments and angular momenta. There are no adjustable parameters in the model and agreement with experiment is very good as shown in Table 3.  
\begin{table}[ht]
\begin{center}
\caption{Properties of the gravitationally confined three-neutrino states}\par
(using $m_o=0.0437\;eV/c^2$ \cite{Vayenas12} for the neutrino mass)
\begin{tabular}{|p{1.7in}|p{1.4in}|p{1.3in}|} \hline 
\textbf{Property} & \textbf{Model predicted value} & \textbf{Experimental value}\\ \hline\hline
\vspace{0.1cm}Neutron rest mass & \vspace{0.1cm} 935.565 $MeV/c^2$ & \vspace{0.1cm}935.565 $MeV/c^2$\\ \hline
\vspace{0.1cm}Baryon binding energy & \vspace{0.1cm} 208 $MeV$ & \vspace{0.1cm} $\sim$150 $MeV$$^*$\\ \hline
\vspace{0.1cm}Reduced de Broglie wavelength\par or radius of ground state & \vspace{0.1cm} 0.631 $fm$ & \vspace{0.1cm} $\sim$0.7 $fm$\\ \hline
\vspace{0.1cm}Minimum lifetime & \vspace{0.1cm} 6.6$\times 10^{-24}$ $s$ & \vspace{0.1cm} 5.6$\times 10^{-24}$ $s$\\ \hline
\vspace{0.1cm}Proton magnetic moment & \vspace{0.1cm} 15.14$\cdot 10^{-27}$ $J/T$ & \vspace{0.1cm}14.10$\cdot 10^{-27}$ $J/T$\\ \hline
Neutron magnetic moment & \vspace{0.1cm} -10.09$\cdot 10^{-27}$ $J/T$ & \vspace{0.1cm} -9.66$\cdot 10^{-27}$ $J/T$\\ \hline
\vspace{0.1cm}Gravitational mass $\gamma^3m_o$ & \vspace{0.1cm} 1.607$\cdot 10^{19}$ $GeV/c^2$ & \vspace{0.1cm} 1.221$\cdot 10^{19}$ $GeV/c^2$\par (Planck mass)\\
\hline
\vspace{0.1cm}Angular momentum & \vspace{0.1cm} 1.13 $(m/GeV)^2$ $\hbar$ $\sim \hbar$ & \vspace{0.1cm} $\sim \hbar$ $\;^{**}$\\ \hline
\end{tabular}
\label{tab:3}
\end{center}
{*} : QCD predicted value at the QCD transition temperature \cite{Braun07}\\ 
{**} : Ref. \cite{Hooft07}\\ 
\end{table}

It thus appears that the large relativistic inertial and gravitational mass of fast neutrinos makes them suitable for constructing much heavier composite particles, such as hadrons. That baryons and mesons may consist of neutrinos is not too surprising since, for example, the $\pi$ and $K$ meson decay produces neutrinos \cite{Griffiths08} and also since neutrinos are known to be emitted in practically all hadron decays and nuclear reactions \cite{Griffiths08,Povh06}. Since neutrinos and antineutrinos come in three flavors with different masses, it appears worthwhile  to test in the future the usefulness of such deterministic Bohr-de Broglie type models using various neutrino and antineutrino combinations and gravity as the attractive force for the possible description of the formation of other composite particles too. 

\section*{Acknowledgements}
CGV acknowledges numerous helpful discussions with Professor Ilan Riess of the Physics Department at the Technion.

\renewcommand{\theequation}{A\arabic{equation}}
\setcounter{equation}{0} 
\section*{APPENDIX A}
Here we present two examples showing the type of paradoxes created if one assumes that the inertial mass for circular motion is $\gamma m_o$.

\textbf{Example 1:} Refering to Fig. 2, which shows via dashed lines that the particle under consideration may be performing a linear or circular motion, one notes that  if the inertial mass of the particle is assumed to equal $\gamma m_o$ for circular motion, then the gravitational mass $m_g=m_i=\gamma m_o$ is different from that $(\gamma^3m_o)$ for linear motion of the same particle with the same rest mass and the same velocity. 

This would imply that the magnitude of the gravitational mass and thus gravitational force exerted between the particle under consideration and any other particle  is not uniquely defined, i.e. has two possible values.

\textbf{Example 2:} Refering to Figure 7 one expects the force $F_G$ between particles A and B to be the same in Figures 7a and 7b since they are at the same distance and have the same velocities. 
\begin{figure}[h]
\begin{center}
\includegraphics[width=0.60\textwidth]{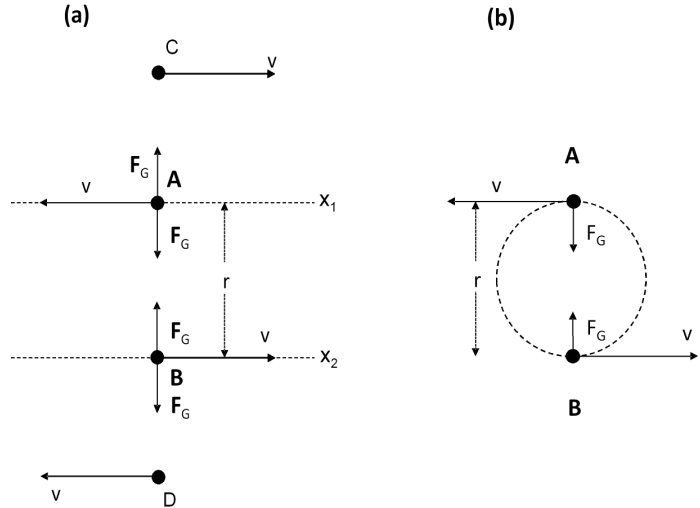}
\caption{\textbf{(a):} Particles A and B moving in parallel lines with constant velocities $\texttt{v}$ exerting to each other a gravitational force $F_G$ which is counterbalanced by attractive forces from two other particles C and D so that particles A and B move on the parallel straight lines $X_1$ and $X_2$ despite their gravitational attraction. \textbf{(b):} Cyclic motion with the same distance $r$, and the same velocity $\texttt{v}$ so that the force between the particles has to be the same as in \textbf{(a)}.}
\label{fig:7}
\end{center}
\end{figure}

Since particles A and B in Fig. 7a move on linear paths it follows: 
\begin{equation}
m_g=m_i=\gamma^3m_o
\label{eq:A1}
\end{equation}
and 
\begin{equation}
F_G=G\frac{m^2_g}{r^2}=G\frac{m^2_o\gamma^6}{r^2}
\label{eq:A2}
\end{equation}
 
However if one erroneously assumes $m_i=\gamma m_o$ for the cyclic motion (Fig. 7b) then it is: 
\begin{equation}
F_G=G\frac{m^2_o\gamma^2}{r^2}
\label{eq:A3}
\end{equation}
which differs from the correct value obtained from eq. (\ref{eq:A2}) by a factor of $\gamma^4$.

As a final note it is worth examining what happens to the present model if one uses the relativistic mass, $\gamma m_o$, in the place of the inertial and gravitational mass $\gamma^3m_o$ in Newton's universal gravitational law. In this case one finds again that rotational bound states can form with the masses and properties of baryons. The masses however of the constituent particles are in this case much smaller than those of neutrinos.

\end{document}